\documentclass[conference]{IEEEtran}
\IEEEoverridecommandlockouts
\usepackage[nolist]{acronym}%
\usepackage{amsmath, amssymb, graphicx}
\usepackage{float, stfloats}
\usepackage{hyperref}
\usepackage[detect-weight,retain-explicit-plus]{siunitx}
\usepackage{subfig}
\usepackage[dvipsnames]{xcolor}
\usepackage{textcomp}
\usepackage{enumerate}

\usepackage{tikz}

\usetikzlibrary{arrows.meta}

\makeatletter

\pgfdeclarearrow{
  name = ipe _linear,
  defaults = {
    length = +1bp,
    width  = +.666bp,
    line width = +0pt 1,
  },
  setup code = {
    \pgfarrowssetbackend{0pt}
    \pgfarrowssetvisualbackend{
      \pgfarrowlength\advance\pgf@x by-.5\pgfarrowlinewidth}
    \pgfarrowssetlineend{\pgfarrowlength}
    \ifpgfarrowreversed
      \pgfarrowssetlineend{\pgfarrowlength\advance\pgf@x by-.5\pgfarrowlinewidth}
    \fi
    \pgfarrowssettipend{\pgfarrowlength}
    \pgfarrowshullpoint{\pgfarrowlength}{0pt}
    \pgfarrowsupperhullpoint{0pt}{.5\pgfarrowwidth}
    \pgfarrowssavethe\pgfarrowlinewidth
    \pgfarrowssavethe\pgfarrowlength
    \pgfarrowssavethe\pgfarrowwidth
  },
  drawing code = {
    \pgfsetdash{}{+0pt}
    \ifdim\pgfarrowlinewidth=\pgflinewidth\else\pgfsetlinewidth{+\pgfarrowlinewidth}\fi
    \pgfpathmoveto{\pgfqpoint{0pt}{.5\pgfarrowwidth}}
    \pgfpathlineto{\pgfqpoint{\pgfarrowlength}{0pt}}
    \pgfpathlineto{\pgfqpoint{0pt}{-.5\pgfarrowwidth}}
    \pgfusepathqstroke
  },
  parameters = {
    \the\pgfarrowlinewidth,%
    \the\pgfarrowlength,%
    \the\pgfarrowwidth,%
  },
}

\pgfdeclarearrow{
  name = ipe _pointed,
  defaults = {
    length = +1bp,
    width  = +.666bp,
    inset  = +.2bp,
    line width = +0pt 1,
  },
  setup code = {
    \pgfarrowssetbackend{0pt}
    \pgfarrowssetvisualbackend{\pgfarrowinset}
    \pgfarrowssetlineend{\pgfarrowinset}
    \ifpgfarrowreversed
      \pgfarrowssetlineend{\pgfarrowlength}
    \fi
    \pgfarrowssettipend{\pgfarrowlength}
    \pgfarrowshullpoint{\pgfarrowlength}{0pt}
    \pgfarrowsupperhullpoint{0pt}{.5\pgfarrowwidth}
    \pgfarrowshullpoint{\pgfarrowinset}{0pt}
    \pgfarrowssavethe\pgfarrowinset
    \pgfarrowssavethe\pgfarrowlinewidth
    \pgfarrowssavethe\pgfarrowlength
    \pgfarrowssavethe\pgfarrowwidth
  },
  drawing code = {
    \pgfsetdash{}{+0pt}
    \ifdim\pgfarrowlinewidth=\pgflinewidth\else\pgfsetlinewidth{+\pgfarrowlinewidth}\fi
    \pgfpathmoveto{\pgfqpoint{\pgfarrowlength}{0pt}}
    \pgfpathlineto{\pgfqpoint{0pt}{.5\pgfarrowwidth}}
    \pgfpathlineto{\pgfqpoint{\pgfarrowinset}{0pt}}
    \pgfpathlineto{\pgfqpoint{0pt}{-.5\pgfarrowwidth}}
    \pgfpathclose
    \ifpgfarrowopen
      \pgfusepathqstroke
    \else
      \ifdim\pgfarrowlinewidth>0pt\pgfusepathqfillstroke\else\pgfusepathqfill\fi
    \fi
  },
  parameters = {
    \the\pgfarrowlinewidth,%
    \the\pgfarrowlength,%
    \the\pgfarrowwidth,%
    \the\pgfarrowinset,%
    \ifpgfarrowopen o\fi%
  },
}

\newdimen\ipeminipagewidth

\tikzstyle{ipe import} = [
  x=1bp, y=1bp,
  ipe node stretch/.store in=\ipenodestretch,
  ipe stretch normal/.style={ipe node stretch=1},
  ipe stretch normal,
  ipe node/.style={
    anchor=base west, inner sep=0, outer sep=0, scale=\ipenodestretch
  },
  ipe mark scale/.store in=\ipemarkscale,
  ipe mark tiny/.style={ipe mark scale=1.1},
  ipe mark small/.style={ipe mark scale=2},
  ipe mark normal/.style={ipe mark scale=3},
  ipe mark large/.style={ipe mark scale=5},
  ipe mark normal, 
  ipe circle/.pic={
    \draw[line width=0.2*\ipemarkscale]
      (0,0) circle[radius=0.5*\ipemarkscale];
    \coordinate () at (0,0);
  },
  ipe disk/.pic={
    \fill (0,0) circle[radius=0.6*\ipemarkscale];
    \coordinate () at (0,0);
  },
  ipe fdisk/.pic={
    \filldraw[line width=0.2*\ipemarkscale]
      (0,0) circle[radius=0.5*\ipemarkscale];
    \coordinate () at (0,0);
  },
  ipe box/.pic={
    \draw[line width=0.2*\ipemarkscale, line join=miter]
      (-.5*\ipemarkscale,-.5*\ipemarkscale) rectangle
      ( .5*\ipemarkscale, .5*\ipemarkscale);
    \coordinate () at (0,0);
  },
  ipe square/.pic={
    \fill
      (-.6*\ipemarkscale,-.6*\ipemarkscale) rectangle
      ( .6*\ipemarkscale, .6*\ipemarkscale);
    \coordinate () at (0,0);
  },
  ipe fsquare/.pic={
    \filldraw[line width=0.2*\ipemarkscale, line join=miter]
      (-.5*\ipemarkscale,-.5*\ipemarkscale) rectangle
      ( .5*\ipemarkscale, .5*\ipemarkscale);
    \coordinate () at (0,0);
  },
  ipe cross/.pic={
    \draw[line width=0.2*\ipemarkscale, line cap=butt]
      (-.5*\ipemarkscale,-.5*\ipemarkscale) --
      ( .5*\ipemarkscale, .5*\ipemarkscale)
      (-.5*\ipemarkscale, .5*\ipemarkscale) --
      ( .5*\ipemarkscale,-.5*\ipemarkscale);
    \coordinate () at (0,0);
  },
  /pgf/arrow keys/.cd,
  ipe arrow normal/.style={scale=1},
  ipe arrow tiny/.style={scale=.4},
  ipe arrow small/.style={scale=.7},
  ipe arrow large/.style={scale=1.4},
  ipe arrow normal,
  /tikz/.cd,
  ipe arrows/.style={
    ipe normal/.tip={
      ipe _pointed[length=1bp, width=.666bp, inset=0bp,
                   quick, ipe arrow normal]},
    ipe pointed/.tip={
      ipe _pointed[length=1bp, width=.666bp, inset=0.2bp,
                   quick, ipe arrow normal]},
    ipe linear/.tip={
      ipe _linear[length = 1bp, width=.666bp,
                  ipe arrow normal, quick]},
    ipe fnormal/.tip={ipe normal[fill=white]},
    ipe fpointed/.tip={ipe pointed[fill=white]},
    ipe double/.tip={ipe normal[] ipe normal},
    ipe fdouble/.tip={ipe fnormal[] ipe fnormal},
    ipe arc/.tip={ipe normal},
    ipe farc/.tip={ipe fnormal},
    ipe ptarc/.tip={ipe pointed},
    ipe fptarc/.tip={ipe fpointed},
  },
  ipe arrows, 
]

\tikzset{
  rgb color/.code args={#1=#2}{%
    \definecolor{tempcolor-#1}{rgb}{#2}%
    \tikzset{#1=tempcolor-#1}%
  },
}

\makeatother

\usetikzlibrary{arrows.meta,patterns}
\usetikzlibrary{ipe} 

\usepackage{makecell}
\usepackage{multirow}
\usepackage{balance}
\usepackage{booktabs}
\usepackage{fontawesome5}

\usepackage[style=ieee, doi=false, isbn=false, url=false, eprint=false, related=false, maxnames=6, minnames=1, date=year]{biblatex}
\AtEveryBibitem{\clearlist{language}}
\AtEveryBibitem{\clearfield{note}}
\AtEveryBibitem{\clearfield{address}}
\AtEveryBibitem{\clearlist{address}}
\AtEveryBibitem{\clearlist{location}}
\AtEveryBibitem{\clearfield{location}}
\AtEveryBibitem{\clearfield{publisher}}
\AtEveryBibitem{\clearlist{publisher}}
\AtEveryBibitem{\clearlist{pages}}
\AtEveryBibitem{\clearfield{pages}}
\AtEveryBibitem{\clearfield{pages}}

\graphicspath{{./Figures}}

\newcommand{\ra}{$\rightarrow$ }
\newcommand{\ua}{$\uparrow$}
\newcommand{\da}{$\downarrow$}

\newcommand{\numcirc}[1]{\raisebox{.5pt}{\textcircled{\raisebox{-.9pt} {#1}}}}

\def\BibTeX{{\rm B\kern-.05em{\sc i\kern-.025em b}\kern-.08em
    T\kern-.1667em\lower.7ex\hbox{E}\kern-.125emX}}

\begin{document}
\title{Deep Learning-based F0 Synthesis for \\Speaker Anonymization
\thanks{The International Audio Laboratories Erlangen are a joint institution of the Friedrich-Alexander-Universität Erlangen-Nürnberg and Fraunhofer IIS.} 
}
\author{\IEEEauthorblockN{Ünal Ege Gaznepoglu, Nils Peters}
\IEEEauthorblockA{\textit{International Audio Laboratories, Friedrich-Alexander-Universität, Am Wolfsmantel 33, 91058 Erlangen, Germany} \\
\{ege.gaznepoglu, nils.peters\}@audiolabs-erlangen.de}
}
\maketitle
\newacro{BN}{bottleneck feature}
\newacro{DNN}{deep neural network}
\newacro{F0}{fundamental frequency}
\newacro{EER}{equal error rate}
\newacro{WER}{word error rate}
\newacro{PI}{personal information}
\newacro{MSE}{mean-squared error}
\newacro{NSF}{neural source-filter}
\newacro{AM}{acoustic model}
\newacro{VPC}{VoicePrivacy Challenge}
\newacro{ASV}{automated speaker verification}
\newacro{ASR}{automated speech recognition}
\newacro{TDNN}{time delay neural network}
\newacro{FPE}{Fine Pitch Error}
\newacro{GPE}{Gross Pitch Error}
\begin{abstract}
Voice conversion for speaker anonymization is an emerging concept for privacy protection. In a deep learning setting, this is achieved by extracting multiple features from speech, altering the speaker identity, and waveform synthesis. However, many existing systems do not modify \ac{F0} trajectories, which convey prosody information and can reveal speaker identity. Moreover, mismatch between F0 and other features can degrade speech quality and intelligibility. In this paper, we formally introduce a method that synthesizes F0 trajectories from other speech features and evaluate its reconstructional capabilities. Then we test our approach within a speaker anonymization framework, comparing it to a baseline and a state-of-the-art F0 modification that utilizes speaker information. The results show that our method improves both speaker anonymity, measured by the \acl{EER}, and utility, measured by the \acl{WER}. 
\end{abstract}
\acresetall
\section{Introduction} \label{sec:intro}

\noindent Speaker anonymization is a feasible solution to prevent \acl{PI} leakage during cloud-enabled speech processing tasks, such as voice assistant usage \cite{tomashenko_introducing_2020}. The VoicePrivacy Initiative organizes a \ac{VPC} to facilitate further studies \cite{tomashenko_vpc_evalplan_2022}. Many researchers adopt their datasets, baselines and evaluation methodology \cite{turner_speaker_2020, champion_study_2021, gaznepoglu_exploring_2021, mawalim_speaker_2022}.

\begin{figure}[t]
    \centering
    \subfloat[\ac{VPC} 2022 Baseline B1.a/b \cite{tomashenko_vpc_evalplan_2022} \label{fig:baseline}] 
    {\resizebox{\linewidth}{!}{\begingroup
\renewcommand{\baselinestretch}{1} \endlinechar=-1 \tikzstyle{ipe stylesheet} = [
  ipe import,
  even odd rule,
  line join=round,
  line cap=butt,
  ipe pen normal/.style={line width=0.4},
  ipe pen heavier/.style={line width=0.8},
  ipe pen fat/.style={line width=1.2},
  ipe pen ultrafat/.style={line width=2},
  ipe pen normal,
  ipe mark normal/.style={ipe mark scale=3},
  ipe mark large/.style={ipe mark scale=5},
  ipe mark small/.style={ipe mark scale=2},
  ipe mark tiny/.style={ipe mark scale=1.1},
  ipe mark normal,
  /pgf/arrow keys/.cd,
  ipe arrow normal/.style={scale=7},
  ipe arrow large/.style={scale=10},
  ipe arrow small/.style={scale=5},
  ipe arrow tiny/.style={scale=3},
  ipe arrow normal,
  /tikz/.cd,
  ipe arrows, 
  <->/.tip = ipe normal,
  ipe dash normal/.style={dash pattern=},
  ipe dash dotted/.style={dash pattern=on 1bp off 3bp},
  ipe dash dashed/.style={dash pattern=on 4bp off 4bp},
  ipe dash dash dotted/.style={dash pattern=on 4bp off 2bp on 1bp off 2bp},
  ipe dash dash dot dotted/.style={dash pattern=on 4bp off 2bp on 1bp off 2bp on 1bp off 2bp},
  ipe dash normal,
  ipe node/.append style={font=\normalsize},
  ipe stretch normal/.style={ipe node stretch=1},
  ipe stretch normal,
  ipe opacity 10/.style={opacity=0.1},
  ipe opacity 30/.style={opacity=0.3},
  ipe opacity 50/.style={opacity=0.5},
  ipe opacity 75/.style={opacity=0.75},
  ipe opacity opaque/.style={opacity=1},
  ipe opacity opaque,
]
\definecolor{red}{rgb}{1,0,0}
\definecolor{blue}{rgb}{0,0,1}
\definecolor{green}{rgb}{0,1,0}
\definecolor{yellow}{rgb}{1,1,0}
\definecolor{orange}{rgb}{1,0.647,0}
\definecolor{gold}{rgb}{1,0.843,0}
\definecolor{purple}{rgb}{0.627,0.125,0.941}
\definecolor{gray}{rgb}{0.745,0.745,0.745}
\definecolor{brown}{rgb}{0.647,0.165,0.165}
\definecolor{navy}{rgb}{0,0,0.502}
\definecolor{pink}{rgb}{1,0.753,0.796}
\definecolor{seagreen}{rgb}{0.18,0.545,0.341}
\definecolor{turquoise}{rgb}{0.251,0.878,0.816}
\definecolor{violet}{rgb}{0.933,0.51,0.933}
\definecolor{darkblue}{rgb}{0,0,0.545}
\definecolor{darkcyan}{rgb}{0,0.545,0.545}
\definecolor{darkgray}{rgb}{0.663,0.663,0.663}
\definecolor{darkgreen}{rgb}{0,0.392,0}
\definecolor{darkmagenta}{rgb}{0.545,0,0.545}
\definecolor{darkorange}{rgb}{1,0.549,0}
\definecolor{darkred}{rgb}{0.545,0,0}
\definecolor{lightblue}{rgb}{0.678,0.847,0.902}
\definecolor{lightcyan}{rgb}{0.878,1,1}
\definecolor{lightgray}{rgb}{0.827,0.827,0.827}
\definecolor{lightgreen}{rgb}{0.565,0.933,0.565}
\definecolor{lightyellow}{rgb}{1,1,0.878}
\definecolor{black}{rgb}{0,0,0}
\definecolor{white}{rgb}{1,1,1}
\begin{tikzpicture}[ipe stylesheet]
  \draw[shift={(104, 768)}, xscale=0.6875, yscale=0.4]
    (0, 0) rectangle (64, -160);
  \draw[shift={(220, 700)}, yscale=0.4, -{>[ipe arrow small]}]
    (0, 0)
     -- (0, 20);
  \draw[shift={(64, 736)}, xscale=0.25, -{>[ipe arrow small]}]
    (0, 0)
     -- (160, 0);
  \node[ipe node, anchor=south west]
     at (70.28, 737) {Input};
  \node[ipe node, anchor=north west]
     at (67.374, 735) {Speech};
  \draw[-{>[ipe arrow small]}]
    (248, 716)
     -- (260, 716)
     -- (260, 724)
     -- (272, 724);
  \draw[-{>[ipe arrow small]}]
    (148, 736)
     -- (272, 736);
  \draw[-{>[ipe arrow small]}]
    (148, 760)
     -- (260, 760)
     -- (260, 748)
     -- (272, 748);
  \node[ipe node, anchor=center, font=\scriptsize]
     at (160, 740) {BN};
  \node[ipe node, anchor=center, font=\scriptsize]
     at (160, 764) {F0};
  \node[ipe node, anchor=west, font=\scriptsize]
     at (260, 708) {anonymized};
  \node[ipe node, anchor=center]
     at (126, 736) {ASR};
  \draw[shift={(108, 744.001)}, xscale=0.75, yscale=0.6667]
    (0, 0) rectangle (48, -24);
  \draw[shift={(108, 724)}, xscale=0.75, yscale=0.6667]
    (0, 0) rectangle (48, -24);
  \node[ipe node, anchor=center]
     at (126, 716) {ASV};
  \begin{scope}[shift={(0, 4)}]
    \draw[-{>[ipe arrow small]}]
      (352, 732)
       -- (392, 732);
    \node[ipe node, anchor=south west]
       at (353.97, 733) {Output};
    \node[ipe node, anchor=north west]
       at (355.077, 731) {Speech};
  \end{scope}
  \draw[shift={(272, 756.002)}, scale=1.6667]
    (0, 0) rectangle (48, -24);
  \node[ipe node, anchor=center]
     at (312, 748) {Neural Vocoder};
  \node[ipe node, anchor=center, font=\small]
     at (312, 736) {(AM-NSF /};
  \node[ipe node, anchor=center, font=\small]
     at (312, 724) {HiFi GAN etc.)};
  \node[ipe node, anchor=west, font=\scriptsize]
     at (260, 700) {x-vector};
  \node[ipe node, anchor=center, font=\small]
     at (126, 696) {Feature};
  \node[ipe node, anchor=center, font=\small]
     at (126, 688) {Extractors};
  \draw[shift={(191.999, 724)}, xscale=1.1667, yscale=0.6667]
    (0, 0) rectangle (48, -24);
  \node[ipe node, anchor=center]
     at (220, 716) {Anonymize};
  \begin{scope}[shift={(0, -4)}]
    \draw[shift={(108, 768)}, xscale=0.75, yscale=0.6667]
      (0, 0) rectangle (48, -24);
    \node[ipe node, anchor=center]
       at (126, 760) {F0 Ext.};
  \end{scope}
  \begin{scope}[shift={(0, 8)}]
    \node[ipe node, anchor=center, font=\small]
       at (224, 684) {X-vector pool};
    \node[ipe node, anchor=center, font=\scriptsize]
       at (192, 684) {\faDatabase};
    \draw[shift={(184, 692)}, yscale=0.6667]
      (0, 0) rectangle (72, -24);
  \end{scope}
  \begin{scope}[shift={(0, 4)}]
    \draw[shift={(148, 712)}, xscale=0.9167, -{>[ipe arrow small]}]
      (0, 0)
       -- (48, 0);
    \node[ipe node, anchor=center, font=\scriptsize]
       at (168, 716) {X-vector};
  \end{scope}
\end{tikzpicture}\endgroup \renewcommand{\baselinestretch}{1.5}}}
    \\
    \subfloat[State-of-the-art F0 modification proposed by \cite{champion_study_2021} \label{fig:Champion}]
    {\resizebox{\linewidth}{!}{\begingroup
\renewcommand{\baselinestretch}{1} \endlinechar=-1 \tikzstyle{ipe stylesheet} = [
  ipe import,
  even odd rule,
  line join=round,
  line cap=butt,
  ipe pen normal/.style={line width=0.4},
  ipe pen heavier/.style={line width=0.8},
  ipe pen fat/.style={line width=1.2},
  ipe pen ultrafat/.style={line width=2},
  ipe pen normal,
  ipe mark normal/.style={ipe mark scale=3},
  ipe mark large/.style={ipe mark scale=5},
  ipe mark small/.style={ipe mark scale=2},
  ipe mark tiny/.style={ipe mark scale=1.1},
  ipe mark normal,
  /pgf/arrow keys/.cd,
  ipe arrow normal/.style={scale=7},
  ipe arrow large/.style={scale=10},
  ipe arrow small/.style={scale=5},
  ipe arrow tiny/.style={scale=3},
  ipe arrow normal,
  /tikz/.cd,
  ipe arrows, 
  <->/.tip = ipe normal,
  ipe dash normal/.style={dash pattern=},
  ipe dash dotted/.style={dash pattern=on 1bp off 3bp},
  ipe dash dashed/.style={dash pattern=on 4bp off 4bp},
  ipe dash dash dotted/.style={dash pattern=on 4bp off 2bp on 1bp off 2bp},
  ipe dash dash dot dotted/.style={dash pattern=on 4bp off 2bp on 1bp off 2bp on 1bp off 2bp},
  ipe dash normal,
  ipe node/.append style={font=\normalsize},
  ipe stretch normal/.style={ipe node stretch=1},
  ipe stretch normal,
  ipe opacity 10/.style={opacity=0.1},
  ipe opacity 30/.style={opacity=0.3},
  ipe opacity 50/.style={opacity=0.5},
  ipe opacity 75/.style={opacity=0.75},
  ipe opacity opaque/.style={opacity=1},
  ipe opacity opaque,
]
\definecolor{red}{rgb}{1,0,0}
\definecolor{blue}{rgb}{0,0,1}
\definecolor{green}{rgb}{0,1,0}
\definecolor{yellow}{rgb}{1,1,0}
\definecolor{orange}{rgb}{1,0.647,0}
\definecolor{gold}{rgb}{1,0.843,0}
\definecolor{purple}{rgb}{0.627,0.125,0.941}
\definecolor{gray}{rgb}{0.745,0.745,0.745}
\definecolor{brown}{rgb}{0.647,0.165,0.165}
\definecolor{navy}{rgb}{0,0,0.502}
\definecolor{pink}{rgb}{1,0.753,0.796}
\definecolor{seagreen}{rgb}{0.18,0.545,0.341}
\definecolor{turquoise}{rgb}{0.251,0.878,0.816}
\definecolor{violet}{rgb}{0.933,0.51,0.933}
\definecolor{darkblue}{rgb}{0,0,0.545}
\definecolor{darkcyan}{rgb}{0,0.545,0.545}
\definecolor{darkgray}{rgb}{0.663,0.663,0.663}
\definecolor{darkgreen}{rgb}{0,0.392,0}
\definecolor{darkmagenta}{rgb}{0.545,0,0.545}
\definecolor{darkorange}{rgb}{1,0.549,0}
\definecolor{darkred}{rgb}{0.545,0,0}
\definecolor{lightblue}{rgb}{0.678,0.847,0.902}
\definecolor{lightcyan}{rgb}{0.878,1,1}
\definecolor{lightgray}{rgb}{0.827,0.827,0.827}
\definecolor{lightgreen}{rgb}{0.565,0.933,0.565}
\definecolor{lightyellow}{rgb}{1,1,0.878}
\definecolor{black}{rgb}{0,0,0}
\definecolor{white}{rgb}{1,1,1}
\begin{tikzpicture}[ipe stylesheet]
  \draw[shift={(104, 768)}, xscale=0.6875, yscale=0.4]
    (0, 0) rectangle (64, -160);
  \draw[shift={(220, 696)}, yscale=0.6, -{>[ipe arrow small]}]
    (0, 0)
     -- (0, 20);
  \draw[shift={(64, 736)}, xscale=0.25, -{>[ipe arrow small]}]
    (0, 0)
     -- (160, 0);
  \node[ipe node, anchor=south west]
     at (70.28, 737) {Input};
  \node[ipe node, anchor=north west]
     at (67.374, 735) {Speech};
  \draw[shift={(184, 696)}, xscale=1.7778, yscale=0.6667]
    (0, 0) rectangle (72, -24);
  \draw[-{>[ipe arrow small]}]
    (248, 716)
     -- (260, 716)
     -- (260, 724)
     -- (272, 724);
  \draw[-{>[ipe arrow small]}]
    (148, 736)
     -- (272, 736);
  \draw[-{>[ipe arrow small]}]
    (252, 760)
     -- (260, 760)
     -- (260, 748)
     -- (272, 748);
  \node[ipe node, anchor=center, font=\scriptsize]
     at (160, 740) {BN};
  \node[ipe node, anchor=center, font=\scriptsize]
     at (160, 764) {F0};
  \node[ipe node, anchor=west, font=\scriptsize]
     at (260, 708) {anonymized};
  \node[ipe node, anchor=center]
     at (126, 736) {ASR};
  \draw[shift={(108, 744.001)}, xscale=0.75, yscale=0.6667]
    (0, 0) rectangle (48, -24);
  \begin{scope}[shift={(0, 4)}]
    \draw[-{>[ipe arrow small]}]
      (352, 732)
       -- (392, 732);
    \node[ipe node, anchor=south west]
       at (353.97, 733) {Output};
    \node[ipe node, anchor=north west]
       at (355.077, 731) {Speech};
  \end{scope}
  \draw[shift={(272, 756.002)}, scale=1.6667]
    (0, 0) rectangle (48, -24);
  \node[ipe node, anchor=center]
     at (312, 748) {Neural Vocoder};
  \node[ipe node, anchor=center, font=\small]
     at (312, 736) {(AM-NSF /};
  \node[ipe node, anchor=center, font=\small]
     at (312, 724) {HiFi GAN etc.)};
  \node[ipe node, anchor=west, font=\scriptsize]
     at (260, 700) {x-vector};
  \draw[shift={(171.951, 767.998)}, xscale=1.6687, yscale=0.6667, ipe pen fat]
    (0, 0) rectangle (48, -24);
  \node[ipe node, anchor=center]
     at (212, 760) {F0 Modifications};
  \draw[-{>[ipe arrow small]}]
    (148, 760)
     -- (172, 760);
  \draw[-{>[ipe arrow small]}]
    (220, 724)
     -- (220, 732)
     -- (212, 732)
     -- (212, 752);
  \node[ipe node, anchor=west, font=\scriptsize]
     at (216, 744) {F0 stats};
  \node[ipe node, anchor=center, font=\small]
     at (252, 688) {X-vector pool (w/ F0 stats)};
  \node[ipe node, anchor=center, font=\scriptsize]
     at (192, 688) {\faDatabase};
  \begin{scope}[shift={(0, -4)}]
    \draw[shift={(108, 768)}, xscale=0.75, yscale=0.6667]
      (0, 0) rectangle (48, -24);
    \node[ipe node, anchor=center]
       at (126, 760) {F0 Ext.};
  \end{scope}
  \begin{scope}[shift={(0, 4)}]
    \draw[shift={(108, 720)}, xscale=0.75, yscale=0.6667]
      (0, 0) rectangle (48, -24);
    \node[ipe node, anchor=center]
       at (126, 712) {ASV};
  \end{scope}
  \begin{scope}[shift={(0, 4)}]
    \node[ipe node, anchor=center, font=\small]
       at (126, 692) {Feature};
    \node[ipe node, anchor=center, font=\small]
       at (126, 684) {Extractors};
  \end{scope}
  \begin{scope}[shift={(0, 4)}]
    \draw[shift={(148, 712)}, xscale=0.9167, -{>[ipe arrow small]}]
      (0, 0)
       -- (48, 0);
    \node[ipe node, anchor=center, font=\scriptsize]
       at (168, 716) {X-vector};
  \end{scope}
  \begin{scope}[shift={(0, 4)}]
    \draw[shift={(191.999, 720)}, xscale=1.1667, yscale=0.6667]
      (0, 0) rectangle (48, -24);
    \node[ipe node, anchor=center]
       at (220, 712) {Anonymize};
  \end{scope}
\end{tikzpicture}\endgroup \renewcommand{\baselinestretch}{1.5}}}
    \\
    \subfloat[Our F0 synthesis approach \label{fig:ours}]
    {\resizebox{\linewidth}{!}{\begingroup
\renewcommand{\baselinestretch}{1} \endlinechar=-1 \tikzstyle{ipe stylesheet} = [
  ipe import,
  even odd rule,
  line join=round,
  line cap=butt,
  ipe pen normal/.style={line width=0.4},
  ipe pen heavier/.style={line width=0.8},
  ipe pen fat/.style={line width=1.2},
  ipe pen ultrafat/.style={line width=2},
  ipe pen normal,
  ipe mark normal/.style={ipe mark scale=3},
  ipe mark large/.style={ipe mark scale=5},
  ipe mark small/.style={ipe mark scale=2},
  ipe mark tiny/.style={ipe mark scale=1.1},
  ipe mark normal,
  /pgf/arrow keys/.cd,
  ipe arrow normal/.style={scale=7},
  ipe arrow large/.style={scale=10},
  ipe arrow small/.style={scale=5},
  ipe arrow tiny/.style={scale=3},
  ipe arrow normal,
  /tikz/.cd,
  ipe arrows, 
  <->/.tip = ipe normal,
  ipe dash normal/.style={dash pattern=},
  ipe dash dotted/.style={dash pattern=on 1bp off 3bp},
  ipe dash dashed/.style={dash pattern=on 4bp off 4bp},
  ipe dash dash dotted/.style={dash pattern=on 4bp off 2bp on 1bp off 2bp},
  ipe dash dash dot dotted/.style={dash pattern=on 4bp off 2bp on 1bp off 2bp on 1bp off 2bp},
  ipe dash normal,
  ipe node/.append style={font=\normalsize},
  ipe stretch normal/.style={ipe node stretch=1},
  ipe stretch normal,
  ipe opacity 10/.style={opacity=0.1},
  ipe opacity 30/.style={opacity=0.3},
  ipe opacity 50/.style={opacity=0.5},
  ipe opacity 75/.style={opacity=0.75},
  ipe opacity opaque/.style={opacity=1},
  ipe opacity opaque,
]
\definecolor{red}{rgb}{1,0,0}
\definecolor{blue}{rgb}{0,0,1}
\definecolor{green}{rgb}{0,1,0}
\definecolor{yellow}{rgb}{1,1,0}
\definecolor{orange}{rgb}{1,0.647,0}
\definecolor{gold}{rgb}{1,0.843,0}
\definecolor{purple}{rgb}{0.627,0.125,0.941}
\definecolor{gray}{rgb}{0.745,0.745,0.745}
\definecolor{brown}{rgb}{0.647,0.165,0.165}
\definecolor{navy}{rgb}{0,0,0.502}
\definecolor{pink}{rgb}{1,0.753,0.796}
\definecolor{seagreen}{rgb}{0.18,0.545,0.341}
\definecolor{turquoise}{rgb}{0.251,0.878,0.816}
\definecolor{violet}{rgb}{0.933,0.51,0.933}
\definecolor{darkblue}{rgb}{0,0,0.545}
\definecolor{darkcyan}{rgb}{0,0.545,0.545}
\definecolor{darkgray}{rgb}{0.663,0.663,0.663}
\definecolor{darkgreen}{rgb}{0,0.392,0}
\definecolor{darkmagenta}{rgb}{0.545,0,0.545}
\definecolor{darkorange}{rgb}{1,0.549,0}
\definecolor{darkred}{rgb}{0.545,0,0}
\definecolor{lightblue}{rgb}{0.678,0.847,0.902}
\definecolor{lightcyan}{rgb}{0.878,1,1}
\definecolor{lightgray}{rgb}{0.827,0.827,0.827}
\definecolor{lightgreen}{rgb}{0.565,0.933,0.565}
\definecolor{lightyellow}{rgb}{1,1,0.878}
\definecolor{black}{rgb}{0,0,0}
\definecolor{white}{rgb}{1,1,1}
\begin{tikzpicture}[ipe stylesheet]
  \draw[shift={(104, 768)}, xscale=0.6875, yscale=0.4]
    (0, 0) rectangle (64, -160);
  \draw[shift={(220, 700)}, yscale=0.4, -{>[ipe arrow small]}]
    (0, 0)
     -- (0, 20);
  \draw[shift={(64, 736)}, xscale=0.25, -{>[ipe arrow small]}]
    (0, 0)
     -- (160, 0);
  \node[ipe node, anchor=south west]
     at (70.28, 737) {Input};
  \node[ipe node, anchor=north west]
     at (67.374, 735) {Speech};
  \draw[shift={(148, 716)}, xscale=0.9167, -{>[ipe arrow small]}]
    (0, 0)
     -- (48, 0);
  \draw[-{>[ipe arrow small]}]
    (248, 716)
     -- (260, 716)
     -- (260, 724)
     -- (272, 724);
  \draw[-{>[ipe arrow small]}]
    (148, 736)
     -- (272, 736);
  \draw[shift={(108, 764)}, xscale=0.75, yscale=0.6667, gray]
    (0, 0) rectangle (48, -24);
  \node[ipe node, anchor=center, text=gray]
     at (126, 756) {F0 Ext.};
  \node[ipe node, anchor=center, font=\scriptsize]
     at (168, 720) {X-vector};
  \node[ipe node, anchor=center, font=\scriptsize]
     at (160, 740) {BN};
  \node[ipe node, anchor=west, font=\scriptsize]
     at (260, 708) {anonymized};
  \node[ipe node, anchor=center]
     at (126, 736) {ASR};
  \draw[shift={(108, 744.001)}, xscale=0.75, yscale=0.6667]
    (0, 0) rectangle (48, -24);
  \draw[shift={(108, 724)}, xscale=0.75, yscale=0.6667]
    (0, 0) rectangle (48, -24);
  \node[ipe node, anchor=center]
     at (126, 716) {ASV};
  \begin{scope}[shift={(0, 4)}]
    \draw[-{>[ipe arrow small]}]
      (352, 732)
       -- (392, 732);
    \node[ipe node, anchor=south west]
       at (353.97, 733) {Output};
    \node[ipe node, anchor=north west]
       at (355.077, 731) {Speech};
  \end{scope}
  \draw[shift={(272, 756.002)}, scale=1.6667]
    (0, 0) rectangle (48, -24);
  \node[ipe node, anchor=center]
     at (312, 748) {Neural Vocoder};
  \node[ipe node, anchor=center, font=\small]
     at (312, 736) {(AM-NSF /};
  \node[ipe node, anchor=center, font=\small]
     at (312, 724) {HiFi GAN etc.)};
  \node[ipe node, anchor=west, font=\scriptsize]
     at (260, 700) {x-vector};
  \draw[-{>[ipe arrow small]}]
    (252, 760)
     -- (260, 760)
     -- (260, 748)
     -- (272, 748);
  \draw[shift={(188.001, 768)}, xscale=1.3333, yscale=0.6667, ipe pen fat]
    (0, 0) rectangle (48, -24);
  \node[ipe node, anchor=center]
     at (220, 760) {F0 Regressor};
  \draw[-{>[ipe arrow small]}]
    (172, 736)
     -- (172, 764)
     -- (188, 764);
  \draw[-{>[ipe arrow small]}]
    (260, 724)
     -- (260, 728)
     -- (180, 728)
     -- (180, 756)
     -- (188, 756);
  \node[ipe node, anchor=center, font=\small]
     at (126, 696) {Feature};
  \node[ipe node, anchor=center, font=\small]
     at (126, 688) {Extractors};
  \node[ipe node, anchor=center, font=\scriptsize]
     at (260, 764) {F0};
  \node[ipe node, anchor=center, font=\tiny]
     at (172, 736) {$\bullet$};
  \node[ipe node, anchor=center, font=\tiny]
     at (260, 724) {$\bullet$};
  \draw[shift={(191.999, 724)}, xscale=1.1667, yscale=0.6667]
    (0, 0) rectangle (48, -24);
  \node[ipe node, anchor=center]
     at (220, 716) {Anonymize};
  \begin{scope}[shift={(0, 12)}]
    \draw[shift={(184, 688)}, yscale=0.6667]
      (0, 0) rectangle (72, -24);
    \node[ipe node, anchor=center, font=\small]
       at (224, 680) {X-vector pool};
    \node[ipe node, anchor=center, font=\scriptsize]
       at (192, 680) {\faDatabase};
  \end{scope}
\end{tikzpicture}\endgroup \renewcommand{\baselinestretch}{1.5}}}
    \caption{The three speaker anonymization systems under test.}
    \label{fig:baseline_contributions}
\end{figure}
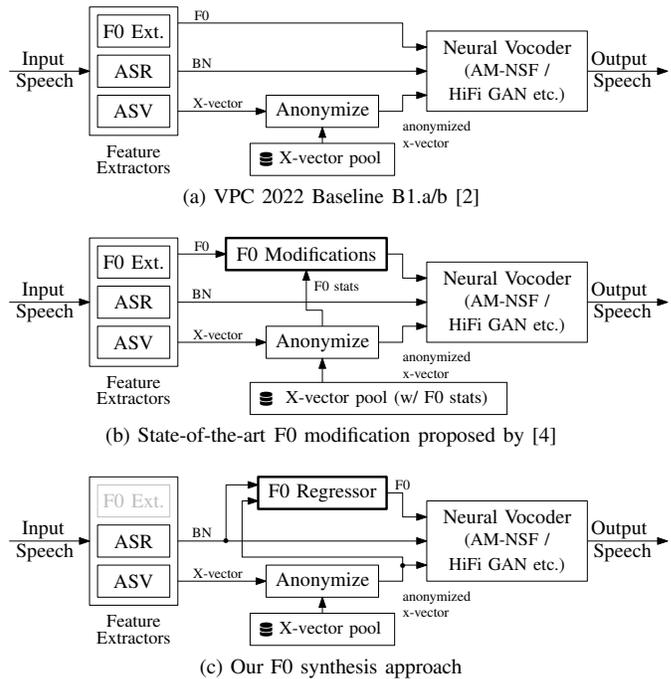

The majority of the studies build upon a system based on X-vectors and neural waveform models \cite{fang_speaker_2019}, citing its superior anonymization performance as well as lower intelligibility degradation, in comparison to the DSP-based anonymization systems \cite{tomashenko_vpc_results_2020}. However, an important caveat of \cite{fang_speaker_2019} is that the \ac{F0} is not altered before synthesis (see Fig.~\ref{fig:baseline}), exposing the original F0, and introducing synthesis artifacts due to the incompatibility between F0 and the anonymized X-vector. Subsequent works \cite{champion_study_2021, gaznepoglu_exploring_2021, tavi_improving_2022} investigated different \ac{F0} manipulations, however, a joint consideration of the extracted features has not been investigated.

In our VPC 2022 contribution \cite{GaznepogluVPC22}, we prototyped a novel approach to address the aforementioned issues simultaneously. Our system, attaining the best naturalness scores in a subjective listening test, was found to be an effective contender \cite{tomashenko_vpc_results_2022}, but an evaluation beyond the VPC paradigm remained out of scope. In this work, we formally test our approach by evaluating its reconstructional capabilities and comparing it with a state-of-the-art F0 modification for anonymization.

\setlength{\textfloatsep}{0.5\baselineskip plus 0.2\baselineskip minus 0.4\baselineskip}

\section{Related Works} \label{sec:literature_review}

\subsection{Speaker anonymization techniques}
\subsubsection{The \texorpdfstring{\ac{VPC}}{} 2022 B1.b Baseline}
The baseline, which our contribution builds upon, is depicted in Fig.~\ref{fig:baseline} \cite{fang_speaker_2019}. It consists of three feature extractors, an anonymization block and a neural vocoder. Extractors decompose speech into individual components, namely  \ac{F0}, \acp{BN}, and X-vectors. Table~\ref{tab:components} summarizes their purposes and extraction details. The anonymization block changes the X-vector, and the neural vocoder (B1.b uses a \ac{NSF} trained with HiFi-GAN discriminators) synthesizes a speech signal from the new set of features. The anonymization block works in the following way:

\begin{enumerate}[(i)]
    \item In the X-vector pool, find a set $\mathcal{P}$ of $N$ X-vectors of a certain gender (same as original or opposite) that are the furthest away (per PLDA) from the original X-vector.
    \item Select $K$ random vectors from $\mathcal{P}$ and average them.
\end{enumerate}

\begin{table}[htbp]
    \centering
    \caption{Extracted features per utterance. W: window size, H: hop size, N: number of frames of an utterance}
    \begin{tabular}{lll}
        \toprule
        \textbf{Feature (purpose)} & \textbf{Extractor} & \textbf{Properties} \\
        \cmidrule{1-1} \cmidrule{2-2} \cmidrule{3-3}
        \acs{F0} (Prosody) & YAAPT \cite{zahorian_spectraltemporal_2008} & size: (Nx1), W: \SI{35}{ms}, H: \SI{10}{ms} \\
        \acs{BN} (Verbal content) & TDNN-F \cite{tomashenko_vpc_evalplan_2022} & size: (Nx256), H: \SI{10}{ms} \\
         X-vector (Identity) & TDNN \cite{tomashenko_vpc_evalplan_2022} & size: (1x512)\\
        \bottomrule
    \end{tabular}
    \label{tab:components}
\end{table}


\subsubsection{Proposed F0 modifications}
Preserving the unmodified \acp{F0} has been identified as a shortcoming, and several improvements were proposed. For instance, a study compared various DSP-based modifications \cite{gaznepoglu_exploring_2021}. Authors of \cite{mawalim_speaker_2022} experimented with speaking rate change and variable \ac{F0} shifting. Notably, Champion et al. \cite{champion_study_2021} proposed creating an F0 statistics dictionary (mean, std) for each speaker in the pool, then assumed the same $K$-subset averages of (mean, std) as the anonymized X-vector statistics. This approach improved the synthesized audio quality, especially for cross-gender anonymization. However, it needs an \ac{F0} dictionary, so it is incompatible with many anonymizers such as \cite{turner_speaker_2020} that have a statistical model instead of a speaker pool. We consider \cite{champion_study_2021}, the only system in the literature that focuses on performing feature-aware F0 modification and reports VPC metrics, as the state-of-the-art, and integrated it into our comparison framework (see Fig.~\ref{fig:Champion}).  

\subsection{Deep learning-based F0 estimation}
\noindent In recent years, data-driven \ac{F0} extractors were developed that outperform the statistics-based ones. CREPE \cite{kim_crepe_2018} and FCN \cite{ardaillon_fully-convolutional_2019} perform a binned classification to estimate \ac{F0}. Tran et al. \cite{tran_robust_2020} uses a joint classification (voiced-unvoiced decision) and regression (\ac{F0} values) formulation. Mentioned systems use convolutional architectures to process the waveforms and are trained on perfectly annotated data. Typical detection metrics such as accuracy, precision and recall are used to evaluate the voiced-unvoiced decision. Fine and gross pitch errors (\acs{FPE}\acused{FPE}, \acs{GPE}\acused{GPE} \cite{vaysse_performance_2022}) are popular metrics to evaluate the \ac{F0} estimation quality. The definitions that we utilized are given below:
\begin{equation}
    \text{GPE: } \frac{\text{num. of frames whose error}>20\%}{\text{num. of correctly identified voiced frames}}
\end{equation}
\begin{equation}
    \text{FPE: } \frac{\text{num. of frames whose error}>5\%}{\text{num. of frames whose error}<20\%}
\end{equation}

\subsection{Speaker anonymization datasets and evaluation} \label{ssec:eval_desc}
\noindent In our work, we adopt the \ac{VPC} framework, consisting of datasets, attack models, and metrics and systems for evaluation. The datasets consist of LibriSpeech and VCTK subsets \cite{tomashenko_vpc_evalplan_2022}. Table~\ref{tab:dataset} outlines our dataset utilization and, which conforms to the \ac{VPC} guidelines. \noindent Further information is available in the \ac{VPC} 2020 \cite{tomashenko_vpc_evalplan_2020} and 2022 \cite{tomashenko_vpc_evalplan_2022} evaluation plans.

\begin{table}[htbp]
    \centering
    \caption{Dataset compositions. \#F, \#M: number of female/male speakers. \#Utt: number of utterances}
    \begin{tabular}{llrrrr}
        \toprule
        \textbf{Used for} & \textbf{Subset Name} & \makecell[c]{\textbf{\#F}} & \makecell[c]{\textbf{\#M}} & \makecell[c]{\textbf{\#Utt}} & \makecell[c]{\textbf{\#Frames}}\\
        \midrule
        \multirow{3}{*}{training} 
        & libri-dev-trials          & 20    & 20    & 1978  & 1411330 \\
        & vctk-dev-trials           & 15    & 15    & 11372 & 3792243 \\
        & libritts-train-clean-100 & 123   & 124   & 33236 & 19297310 \\
        \midrule
        \multirow{2}{*}{validation} 
        & libri-dev-enrolls         & 14    & 15    & 343   & 227416 \\
        & vctk-dev-enrolls          & 15    & 15    & 600   & 192510 \\
        \midrule
        \multirow{2}{*}{testing} 
        & libri-test-\{enrolls,trials\} & 15  & 15    & 1934  & 1563092 \\
        & vctk-test-\{enrolls,trials\}  & 15  & 15    & 12048 & 3846010 \\
        \bottomrule
    \end{tabular}
    \label{tab:dataset}
\end{table}

Anonymization performance is measured using automated speaker verification\acused{ASV}-equal error rate\acused{EER} (\acs{ASV}-\acs{EER}). Higher \acp{EER} correspond to better anonymization as the synthesized speech is less linkable to the original speaker. Some systems introduced in 2020 achieved sufficient anonymization (50\% \ac{EER}) upon comparison to the original speech via a pretrained \ac{ASV} evaluator \cite{tomashenko_vpc_results_2020}. Thus in 2022, a stronger attack model is introduced, where the attackers are able to train \ac{ASV} systems using anonymized data, i.e., a semi-informed attack model \cite{tomashenko_vpc_evalplan_2022}. In our work we use the latter type of \ac{EER} computation.

Moreover, the \ac{WER} in an \ac{ASR} scenario is measured. Lower \acp{WER} are desired, meaning the anonymizer did not compromise the utility. The 2022 challenge trains the \ac{ASR} evaluation system using anonymized data, however the training process is cumbersome, limiting the breadth of our evaluation. Thus we used the 2020 version of \ac{ASR} evaluation which uses a pre-trained model. In 2022, two auxiliary metrics are introduced: pitch correlation ($\rho^{F_0}$) and  gain of voice distinctiveness ($G_\text{VD}$) \cite{noe_towards_2022}. To ensure that the emotions and the speaking pace is largely preserved, $\rho^{F_0}>0.3$ is required. A gain above or below $0$ correspond to an increase or decrease in voice distinctiveness, and $G_\text{VD}=0$ would be the optimal value, attained when the voice distinctiveness is preserved \cite{tomashenko_vpc_evalplan_2022}. 

Some works in the speaker anonymization domain utilize the so-called contrastive systems, that feature minor deviations from the proposed idea, to assess the relative contributions of individual design choices. For instance, in \cite{champion_speaker_2020}, speech files with different feature combinations were synthesized and evaluated, to gain insights on how the acoustic model and the waveform model contribute to the anonymization.

\section{Methodology}
\label{sec:contributions}
\noindent Rather than modifying extracted \ac{F0} trajectories as in previous works, we propose a synthesis by regression approach (see Fig.~\ref{fig:ours}). This approach eliminates the concern of leaking original F0 to the output signal and can potentially improve the quality of the subsequent neural waveform synthesis as the \ac{F0} trajectory would be coherent with the anonymized X-vector.

\subsection{The framewise \texorpdfstring{\ac{F0}}{} synthesizer}

\begin{figure}[b]
    \centering
    \resizebox{\linewidth}{!}{\begingroup
\renewcommand{\baselinestretch}{1} \endlinechar=-1 \tikzstyle{ipe stylesheet} = [
  ipe import,
  even odd rule,
  line join=round,
  line cap=butt,
  ipe pen normal/.style={line width=0.4},
  ipe pen heavier/.style={line width=0.8},
  ipe pen fat/.style={line width=1.2},
  ipe pen ultrafat/.style={line width=2},
  ipe pen normal,
  ipe mark normal/.style={ipe mark scale=3},
  ipe mark large/.style={ipe mark scale=5},
  ipe mark small/.style={ipe mark scale=2},
  ipe mark tiny/.style={ipe mark scale=1.1},
  ipe mark normal,
  /pgf/arrow keys/.cd,
  ipe arrow normal/.style={scale=7},
  ipe arrow large/.style={scale=10},
  ipe arrow small/.style={scale=5},
  ipe arrow tiny/.style={scale=3},
  ipe arrow normal,
  /tikz/.cd,
  ipe arrows, 
  <->/.tip = ipe normal,
  ipe dash normal/.style={dash pattern=},
  ipe dash dotted/.style={dash pattern=on 1bp off 3bp},
  ipe dash dashed/.style={dash pattern=on 4bp off 4bp},
  ipe dash dash dotted/.style={dash pattern=on 4bp off 2bp on 1bp off 2bp},
  ipe dash dash dot dotted/.style={dash pattern=on 4bp off 2bp on 1bp off 2bp on 1bp off 2bp},
  ipe dash normal,
  ipe node/.append style={font=\normalsize},
  ipe stretch normal/.style={ipe node stretch=1},
  ipe stretch normal,
  ipe opacity 10/.style={opacity=0.1},
  ipe opacity 30/.style={opacity=0.3},
  ipe opacity 50/.style={opacity=0.5},
  ipe opacity 75/.style={opacity=0.75},
  ipe opacity opaque/.style={opacity=1},
  ipe opacity opaque,
]

\definecolor{red}{rgb}{1,0,0}
\definecolor{blue}{rgb}{0,0,1}
\definecolor{green}{rgb}{0,1,0}
\definecolor{yellow}{rgb}{1,1,0}
\definecolor{orange}{rgb}{1,0.647,0}
\definecolor{gold}{rgb}{1,0.843,0}
\definecolor{purple}{rgb}{0.627,0.125,0.941}
\definecolor{gray}{rgb}{0.745,0.745,0.745}
\definecolor{brown}{rgb}{0.647,0.165,0.165}
\definecolor{navy}{rgb}{0,0,0.502}
\definecolor{pink}{rgb}{1,0.753,0.796}
\definecolor{seagreen}{rgb}{0.18,0.545,0.341}
\definecolor{turquoise}{rgb}{0.251,0.878,0.816}
\definecolor{violet}{rgb}{0.933,0.51,0.933}
\definecolor{darkblue}{rgb}{0,0,0.545}
\definecolor{darkcyan}{rgb}{0,0.545,0.545}
\definecolor{darkgray}{rgb}{0.663,0.663,0.663}
\definecolor{darkgreen}{rgb}{0,0.392,0}
\definecolor{darkmagenta}{rgb}{0.545,0,0.545}
\definecolor{darkorange}{rgb}{1,0.549,0}
\definecolor{darkred}{rgb}{0.545,0,0}
\definecolor{lightblue}{rgb}{0.678,0.847,0.902}
\definecolor{lightcyan}{rgb}{0.878,1,1}
\definecolor{lightgray}{rgb}{0.827,0.827,0.827}
\definecolor{lightgreen}{rgb}{0.565,0.933,0.565}
\definecolor{lightyellow}{rgb}{1,1,0.878}
\definecolor{black}{rgb}{0,0,0}
\definecolor{white}{rgb}{1,1,1}
\begin{tikzpicture}[ipe stylesheet]
  \draw[shift={(44, 788)}, xscale=0.625, -{>[ipe arrow small]}]
    (0, 0)
     -- (64, 0);
  \draw[shift={(44, 748)}, xscale=0.625, -{>[ipe arrow small]}]
    (0, 0)
     -- (64, 0);
  \node[ipe node, anchor=south east, font=\small]
     at (76, 792) {x-vector};
  \node[ipe node, anchor=south east, font=\small]
     at (76, 752) {BN $[n]$};
  \node[ipe node, anchor=north east, font=\small]
     at (76, 784) {$(1, 512)$};
  \node[ipe node, anchor=north east, font=\small]
     at (76, 744) {$(1, 256)$};
  \draw[shift={(240, 776)}, xscale=0.5, -{>[ipe arrow small]}]
    (0, 0)
     -- (64, 0);
  \node[ipe node, anchor=west, font=\small]
     at (292, 784) {$\hat{F_0}$ $[n]$};
  \draw[-{>[ipe arrow small]}]
    (272, 760)
     -- (280, 760)
     -- (280, 768);
  \draw[shift={(288, 776)}, xscale=1.75, -{>[ipe arrow small]}]
    (0, 0)
     -- (16, 0);
  \begin{scope}[shift={(-120, 40)}]
    \draw
      (400, 736) circle[radius=8];
    \draw
      (396, 740)
       -- (404, 732);
    \draw
      (404, 740)
       -- (396, 732);
  \end{scope}
  \draw[shift={(212, 784)}, xscale=0.6667, yscale=0.2]
    (0, 0) rectangle (48, -160);
  \node[ipe node, anchor=center, font=\small]
     at (220, 776) {FC};
  \node[ipe node, anchor=center, font=\small]
     at (220, 760) {2};
  \draw[shift={(240, 760)}, xscale=0.25, -{>[ipe arrow small]}]
    (0, 0)
     -- (64, 0);
  \node[ipe node]
     at (236, 792) {$f$: {ELU}};
  \node[ipe node, anchor=north, font=\small]
     at (264, 748) {$> 0$};
  \begin{scope}[shift={(-28, 0)}]
    \draw
      (264, 760) circle[radius=4];
    \node[ipe node, anchor=center, font=\footnotesize]
       at (264, 760) {$2$};
  \end{scope}
  \begin{scope}[shift={(-28, 0)}]
    \draw
      (264, 776) circle[radius=4];
    \node[ipe node, anchor=center, font=\footnotesize]
       at (264, 776) {$1$};
  \end{scope}
  \begin{scope}[shift={(-32, 0)}]
    \draw[shift={(288, 768.001)}, yscale=0.6667]
      (0, 0) rectangle (16, -24);
    \draw[shift={(292, 756)}, yscale=0.5]
      (0, 0)
       -- (4, 0)
       -- (4, 16)
       -- (8, 16);
  \end{scope}
  \draw[shift={(148, 792)}, xscale=0.3333, yscale=0.3]
    (0, 0) rectangle (48, -160);
  \node[ipe node, anchor=center, font=\small]
     at (156, 776) {FC};
  \node[ipe node, anchor=center, font=\small]
     at (156, 760) {32};
  \draw[shift={(84, 800)}, xscale=0.3333, yscale=0.4]
    (0, 0) rectangle (48, -160);
  \node[ipe node, anchor=center, font=\small]
     at (92, 776) {FC};
  \node[ipe node, anchor=center, font=\small]
     at (92, 760) {256};
  \draw[shift={(116, 796)}, xscale=0.3333, yscale=0.35]
    (0, 0) rectangle (48, -160);
  \node[ipe node, anchor=center, font=\small]
     at (124, 776) {FC};
  \node[ipe node, anchor=center, font=\small]
     at (124, 760) {64};
  \draw[shift={(180, 788)}, xscale=0.3333, yscale=0.25]
    (0, 0) rectangle (48, -160);
  \node[ipe node, anchor=center, font=\small]
     at (188, 776) {FC};
  \node[ipe node, anchor=center, font=\small]
     at (188, 760) {16};
  \draw[shift={(99.999, 768)}, xscale=0.6667, -{>[ipe arrow small]}]
    (0, 0)
     -- (24, 0);
  \node[ipe node, anchor=center, font=\small]
     at (106.4, 776) {f};
  \draw[shift={(131.999, 768)}, xscale=0.6667, -{>[ipe arrow small]}]
    (0, 0)
     -- (24, 0);
  \node[ipe node, anchor=center, font=\small]
     at (138.4, 776) {f};
  \draw[shift={(163.999, 768)}, xscale=0.6667, -{>[ipe arrow small]}]
    (0, 0)
     -- (24, 0);
  \node[ipe node, anchor=center, font=\small]
     at (170.4, 776) {f};
  \draw[shift={(196, 768)}, xscale=0.6667, -{>[ipe arrow small]}]
    (0, 0)
     -- (24, 0);
  \node[ipe node, anchor=center, font=\small]
     at (202.4, 776) {f};
\end{tikzpicture}\endgroup \renewcommand{\baselinestretch}{1.5}}
    \caption{Proposed architecture. 'FC' is a fully connected layer, with number of neurons indicated below. The output layer neurons are denoted with \numcirc{1}, \numcirc{2}.}
    \label{fig:proposed_nn}
\end{figure}
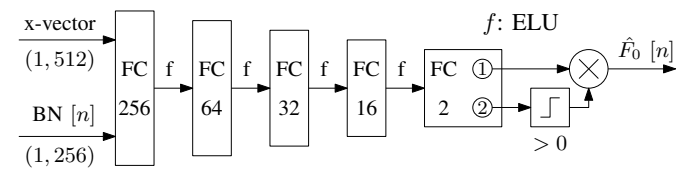

\noindent We propose a 4-hidden-layer \acs{DNN} (see Fig.~\ref{fig:proposed_nn}) to framewise infer F0 from the X-vector and \acp{BN}. Similar to \cite{tran_robust_2020}, it yields two outputs: $\hat{F}_0[n]$, the predicted log-\ac{F0}, and $g[n]$, the logits of a frame being voiced. Then, a mask is constructed, zeroing $\hat{F}_0[n]$ if the frame is classified as unvoiced (i.e., $g<0$). 

\noindent The model is trained on a composite loss function as in \cite{tran_robust_2020}.
\begin{equation}
    \label{eqn:loss} \mathcal{L}(F_0, \hat{F_0}, g, v) = \text{L1}(F_0, \hat{F_0}) + \alpha \text{BCE}(g, v)
\end{equation}  A factor $\alpha$ balances the regression and classification tasks. Here, $v=1$ denotes a voiced frame whereas $v=0$ is an unvoiced one. The functions L1$(\cdot)$ and BCE$(\cdot)$ denote the L1 loss and binary cross-entropy with logits. Different to \cite{tran_robust_2020}, we do not have access to the perfect F0 annotations, so we assume YAAPT extractions as the ground truth. To diminish the effect of imperfect labels, we use L1 loss instead of MSE.

\begin{table}[hbp]
    \centering
    \caption{Hyperparameter search intervals and adopted values. All parameters are searched in the log space.}
    \begin{tabular}{clc}
        \toprule
        \makecell[c]{\textbf{Search Space}}  &  \makecell[l]{\textbf{Purpose}}  &  \makecell[l]{\textbf{Adopted Value}}\\
        \midrule
        $ 10^{-3} < \alpha <  500$                 & loss factor         & $28.112$ \\
        $ 10^{-4} < \delta < 0.5$               & dropout probability & $0$      \\
        N/A                                     & learning rate       & $0.0003$ \\
        N/A                                     & batch size       & $262144$ \\
        \bottomrule
    \end{tabular}\label{tab:hyperparameters}
\end{table}
\vspace{-10pt}
\subsection{Training strategies and hyperparameter tuning}

\noindent The model is implemented with PyTorch \cite{paszke_pytorch_2019}. The training logic is provided by PyTorch Ignite \cite{fomin_high-level_2020}. Training utterances (see Table \ref{tab:dataset}) are concatenated into a single tall matrix and shuffled. This way, voiced and unvoiced frames from different utterances are present in each batch. We used Nesterov-Adam (NAdam) as our optimizer, with default parameters. Early stopping after $10$ epochs, and learning rate reduction by a factor of $0.2$ after $5$ epochs of non-increasing validation metric are used to combat overfitting. We used the total percentage of accurately processed frames, i.e., correctly classified unvoiced frames and frames that do not have gross pitch errors as the validation metric. Table \ref{tab:hyperparameters} outlines the hyperparameters and their tuning procedure with OpTuna \cite{akiba_optuna_2019}. The procedure yielded a $\delta<0.001$, so we disabled dropout by setting $\delta=0$. 

\begin{figure*}[b] 
    \centering
    \begingroup
\renewcommand{\baselinestretch}{1} \endlinechar=-1 \input{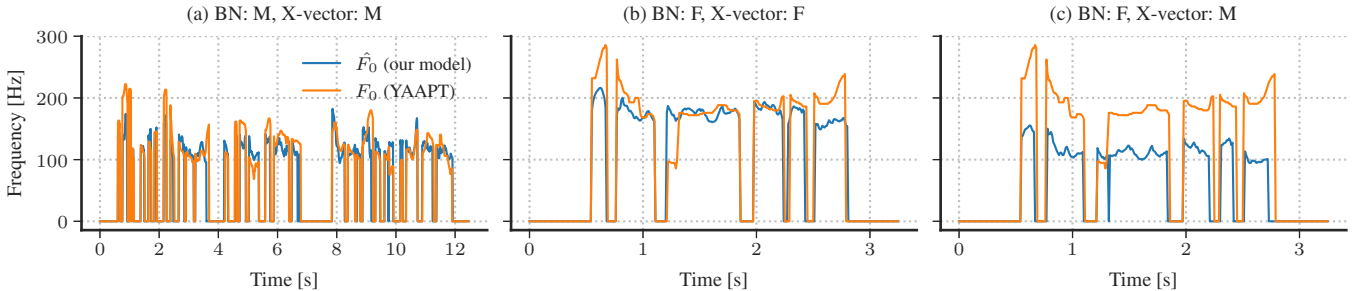}\endgroup \renewcommand{\baselinestretch}{1.5}
    \vspace{-20pt}
    \caption{Model behavior with samples from the validation set. Figures (a) and (b) illustrate reconstructions for those utterances and (c) shows synthesized F0 for the utterance said in (b), if it were said by the speaker in (a).}
    \label{fig:reconstructions}
\end{figure*}

\subsection{Evaluation} \label{sec:evaluation}

\noindent Figures~\ref{fig:reconstructions}a and \ref{fig:reconstructions}b compare the predictions by our model with the assumed ground truth, for a male and a female utterance from the validation set. Fig.~\ref{fig:reconstructions}c shows a cross-gender (female to male) F0 conversion. The predicted F0 (blue) has a mean that is significantly less than the original F0 (orange) while preserving the global trend, hinting that our model is aware of the effects of gender on F0. Now we utilize a two step evaluation procedure to assess the capabilities of our model.

\subsubsection{Reconstruction of known \texorpdfstring{\acp{F0}}{}} \label{sec:reconstruction}
\noindent To measure the similarity of the synthesized F0 values to the assumed ground truth, we compute \ac{GPE}, \ac{FPE} and voiced-unvoiced classification metrics (Accuracy, Precision, Recall).

\begin{figure*}
    \centering
    {{\begingroup
\renewcommand{\baselinestretch}{1} \endlinechar=-1 \input{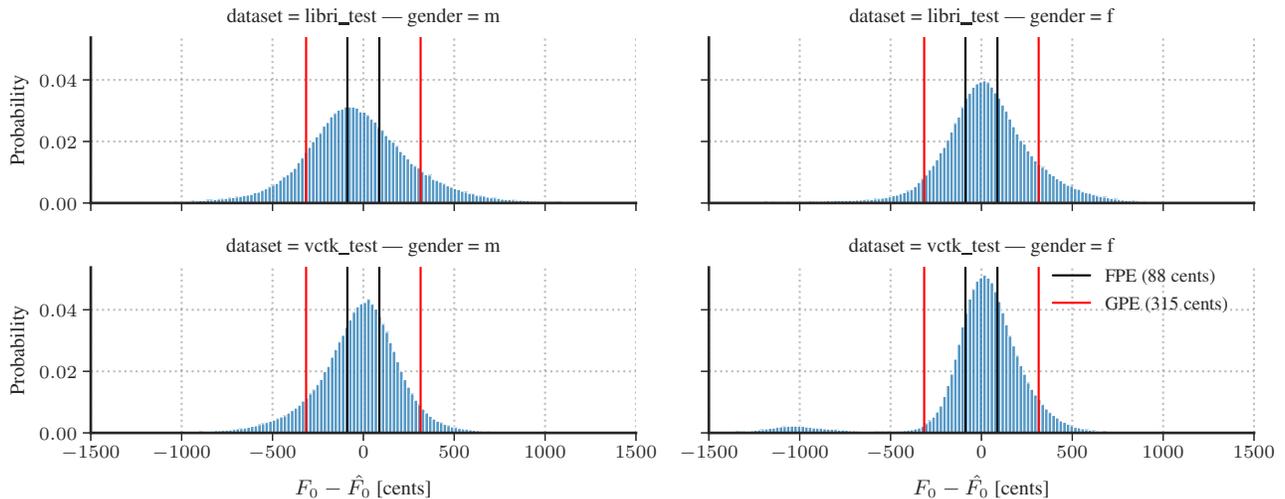}\endgroup \renewcommand{\baselinestretch}{1.5}}}
    \caption{Test set F0 prediction errors. Red and black bars denote the GPE ($315$ cents / 20\%) and FPE ($88$ cents / 5\%) thresholds.}
    \label{fig:reconstruction_errors_cents}
    \vspace{-5pt}
\end{figure*}

\subsubsection{Utilizing the model in a speaker anonymization system}

\noindent We integrate our model into the Baseline B1.b and evaluate according to the \ac{VPC} framework (see Section \ref{ssec:eval_desc}). We also introduce contrastive systems in Table~\ref{tab:contrastive}, which use altered neural vocoder inputs to gain insights. The anonymization block behaves the same across trials for a fair comparison.
\begin{table}[t]
    \centering
    \caption{Contrastive systems and their feature descriptions.}
    \begin{tabular}{llc}
        \toprule
        \textbf{ID}          & \textbf{Input \ac{F0} to \ac{NSF}}   & \textbf{Input X-vector to \ac{NSF}} \\
        \midrule
        Ours                 & synthesized with anonymized X-vector & anonymized \\
        \midrule
        C1                   & synthesized with original X-vector   & original \\
        C2                   & synthesized with anonymized X-vector & original \\
        C3                   & synthesized with original X-vector   & anonymized   \\
        \bottomrule
    \end{tabular}\label{tab:contrastive}
\end{table}

\vspace{-10pt}
\section{Results and Discussion}
\subsection{Reconstruction error of known \texorpdfstring{\acp{F0}}{}} \label{sec:r_reconstruction}

\noindent Table~\ref{tab:results_reconstruction} shows the \ac{F0} reconstruction performance of our model. Despite its simplicity, in particular the narrow temporal scope, it attains a voiced-unvoiced decision test accuracy around $94\%$, similar to the YAAPT's performance reported in \cite{vaysse_performance_2022}. The visual inspection of the synthesized F0 suggests that most decision errors occur at the edges of voiced segments, instead of erratic switchings at random instances. 

The regression performance, measured by GPE and FPE, indicates that our system is only able to provide a crude approximation of the ground truth. End-to-end systems that operate on the full waveform report better GPE and FPE (e.g., \cite{tran_robust_2020}). A histogram of the voiced frame prediction errors is provided in Fig.~\ref{fig:reconstruction_errors_cents}. They resemble a Gaussian distribution with a high variance, centered near zero. We conclude a single frame contains partial information to model the F0 trajectories, and we expect the reconstruction performance to improve if the model used the temporal context, e.g., via a recurrent or convolutional architecture. The lack of perfect F0 annotations also bounds the achievable performance, possibly fixable by a self-supervised training scheme such as SPICE \cite{gfeller_spice_2020}.
\setlength{\tabcolsep}{4pt}

\begin{table}[tb]
\centering
\caption{Model \ac{F0} reconstruction performance, all reported in percentages. Acc: Accuracy, Prec: Precision, Rec: Recall}
\begin{tabular}{l c c c c c c}
    \toprule
    \textbf{Dataset} & \textbf{Sex} & \textbf{GPE}(\da) & \textbf{FPE}(\da) & \textbf{Acc.}(\ua) & \textbf{Prec.}(\ua) & \textbf{Rec.}(\ua) \\
    \cmidrule(lr){1-2} \cmidrule(lr){3-4} \cmidrule(lr){5-7}
    \multirow{2}{*}{{libri-test}}   & F & 31.6 & 66.9 & 93.0 & 94.6 & 93.3 \\
                                    & M & 41.8 & 71.8 & 92.5 & 93.0 & 93.0 \\
    \cmidrule(lr){1-2}              \cmidrule(lr){3-4}   \cmidrule(lr){5-7}
    \multirow{2}{*}{{vctk-test}}    & F & 24.6 & 63.9 & 95.1 & 94.1 & 93.5 \\
                                    & M & 38.8 & 69.9 & 94.6 & 93.5 & 92.5 \\

    \bottomrule
\end{tabular}
\label{tab:results_reconstruction}
\end{table}

\subsection{Utilizing the synthesizer in a speaker anonymization system} \label{sec:r_vpc}

\begin{table*}[!b]
\centering
\vspace{-5pt}
\caption{\ac{VPC} framework results for the baseline (B1.b), our implementation of the state-of-the-art \cite{champion_study_2021}, our proposal, and contrastive systems (C1-C3). Cross gender conversion is only possible for systems utilizing the anonymized X-vector at least once. The weights for the average are introduced by the \ac{VPC} 2022.}
\begin{tabular}{l c c r r r r r r r r r r r r r r}

\toprule
\makecell[c]{\multirow{2}{*}{Dataset}} & 
\multirow{2}{*}{Weight} & Gender &
\multicolumn{6}{c}{EER \cite{tomashenko_vpc_evalplan_2022} [\%] (\ua)} & 
\multicolumn{6}{c}{WER \cite{tomashenko_vpc_evalplan_2020} [\%] (\da)}\\

\cmidrule(lr){4-9} \cmidrule(lr){10-15}
                                 & & (From \ra To) & \makecell[c]{B1.b} & \makecell[c]{\cite{champion_study_2021}} & \makecell[c]{Ours} & \makecell[c]{C1} & \makecell[c]{C2} & \makecell[c]{C3} & \makecell[c]{B1.b} & \makecell[c]{\cite{champion_study_2021}} & \makecell[c]{Ours} & \makecell[c]{C1} & \makecell[c]{C2} & \makecell[c]{C3} \\
\cmidrule(lr){1-3} \cmidrule(lr){4-6} \cmidrule(lr){7-9} \cmidrule(lr){10-12} \cmidrule(lr){13-15}

\multirow{2}{*}{libri-test}     & \multirow{2}{*}{0.25} & F \ra  F & {11.13} & \textbf{12.96}   & {12.04} & 14.60 & 13.14 & 8.57 & \multirow{2}{*}{{5.60}} & \multirow{2}{*}{{5.58}} & \multirow{2}{*}{\textbf{5.48}} & \multirow{2}{*}{6.39} & \multirow{2}{*}{{6.40}} & \multirow{2}{*}{{5.59}}\\ 
                                & & M \ra  M & {7.35} & {8.69}    & \textbf{10.02} & 1.34 & 1.34 & 8.46 & & & & \\
\cmidrule(lr){1-3} \cmidrule(lr){4-6} \cmidrule(lr){7-9} \cmidrule(lr){10-12} \cmidrule(lr){13-15}
\multirow{2}{*}{vctk-test-diff}  & \multirow{2}{*}{0.20} & F \ra  F & {12.04} & {13.73} & \textbf{16.20} & 5.45 & 8.39 & 12.81 & \multirow{4}{*}{{14.66}} & \multirow{4}{*}{{14.76}} & \multirow{4}{*}{\textbf{14.57}} & \multirow{4}{*}{16.28} & \multirow{4}{*}{{16.10}} & \multirow{4}{*}{{14.66}} \\
                                 & & M \ra  M & { 8.78} & {9.93}  & \textbf{10.79} & 1.61 & 2.01 & 8.84 & & & \\
\multirow{2}{*}{vctk-test-com}   &\multirow{2}{*}{0.05} & F \ra  F & {11.56} & {15.03} & \textbf{18.50} &  2.31 & 4.04 & 15.32 & & & \\
                                 & & M \ra  M & { 9.04} & {12.71} & \textbf{14.12} & 1.41 & 0.84 & 11.30 & & & \\
\cmidrule(lr){1-3} \cmidrule(lr){4-6} \cmidrule(lr){7-9} \cmidrule(lr){10-12} \cmidrule(lr){13-15}
\multicolumn{3}{l}{weighted average / same gender} & 9.81 & 11.53 & \textbf{12.54} & 5.58 & 5.94 & 9.92 & 10.13 & 10.17 & \textbf{10.03} & 11.34 & 11.25 & 10.13 \\
\midrule
\multirow{2}{*}{libri-test}      & \multirow{2}{*}{0.25} & F \ra  M &  {14.23} & \textbf{23.18} & {22.99} & N/A & 13.50 & 12.77 & \multirow{2}{*}{{5.99}} & \multirow{2}{*}{{5.82}} & \multirow{2}{*}{\textbf{5.66}} & \multirow{2}{*}{N/A} & \multirow{2}{*}{{6.79}} & \multirow{2}{*}{{5.87}}\\
                                 & & M \ra  F &  {8.46} &  {15.81} & \textbf{19.38} & N/A & 1.34 & 9.8 & & & \\
\cmidrule(lr){1-3} \cmidrule(lr){4-6} \cmidrule(lr){7-9} \cmidrule(lr){10-12} \cmidrule(lr){13-15}
\multirow{2}{*}{vctk-test-diff}  & \multirow{2}{*}{0.20} & F \ra  M & {16.67} &  {26.75} & \textbf{27.83} & N/A & 5.14 & 17.80 & \multirow{4}{*}{{15.37}} & \multirow{4}{*}{{14.98}} & \multirow{4}{*}{\textbf{14.80}} & \multirow{4}{*}{N/A} & \multirow{4}{*}{{17.04}} & \multirow{4}{*}{{15.13}}\\
                                 & & M \ra  F & {14.24} &  {22.62} & \textbf{29.97} & N/A & 2.53 & 22.90 & & & \\
\multirow{2}{*}{vctk-test-com}   & \multirow{2}{*}{0.05} & F \ra  M & {21.39} &  {36.99} & \textbf{38.15} & N/A & 4.36 & 26.88 & & & \\
                                 & & M \ra  F & {12.99} &  {27.97} & \textbf{33.05} & N/A & 1.70 & 18.36 & & & \\
\cmidrule(lr){1-3} \cmidrule(lr){4-6} \cmidrule(lr){7-9} \cmidrule(lr){10-12} \cmidrule(lr){13-15}
\multicolumn{3}{l}{weighted average / cross gender} & 13.57 & 22.87 & \textbf{25.71} & N/A & 5.55 & 16.04 & 10.68 & 10.4 & \textbf{10.23} & N/A & 11.92 & 10.5 \\

\bottomrule
\end{tabular}
\label{tab:results_vpc}
\end{table*}

\noindent Evaluation of the mentioned anonymizers is presented in Table \ref{tab:results_vpc}. The new, stronger attack model caused some decrease in EER for the shift-and-scale approach \cite{champion_study_2021} yet it still outperforms the baseline. The different vocoder resulted in better WERs compared to the original publication, however the conclusions are the same: cross-gender synthesis became more intelligible and same-gender synthesis is comparable to the baseline. Our methodology alters the F0 trajectory altogether, thus on average performs significantly better than the other systems in terms of both metrics, in same gender and cross-gender anonymization. Combined with the observations from the reconstruction performance, in particular the voiced-unvoiced decision differences between YAAPT and our method, we think that our system is possibly able to correct some of the mistakes made by YAAPT thanks to the \acp{BN} and this would explain the WER improvement. 

Evaluation of the contrastive systems provide additional intuition on understanding how F0 modification helps. Usage of the original X-vectors together with the F0 modification (systems C1 and C2) do not yield any significant anonymization and cause an unexpected WER increase (+1\%). We plan to further investigate the reasons for this increase. Supplying the synthesized F0 using the original speaker identity but using the anonymized X-vector for synthesis (system C3) yields an insignificant EER improvement and causes no WER change with respect to the baseline. Hence, it could conceivably be hypothesised that the performance increase yielded by our system is due to the learned characteristics of the speakers and not due to the artifacts our system introduces.

Not mentioned in Table \ref{tab:results_vpc}, our system satisfies the VPC requirement $\rho_{F_0}>0.3$ on all subsets. The $G_{VD}$ values are comparable across the primary systems and indicate a common loss of voice distinctiveness. Previous studies have already shown that the anonymization block is the culprit, because it yields unnaturally similar anonymized X-vectors \cite{turner_speaker_2020}.

Besides the improved metrics, we observe an improved run time for F0 computation and thus speaker anonymization: For all datasets utilized except 'libritts-train-clean-100', it takes only two minutes to synthesize F0 values using our approach, whereas YAAPT extractions take 35 times longer.

\section{Conclusion}

\noindent In this work, we formally evaluated a DNN-based F0 synthesis approach for speaker anonymization. Notwithstanding the architectural simplicity and the lack of perfect F0 annotations for training, the proposed approach managed to improve the \ac{EER} and \ac{WER} metrics over the state-of-the-art speaker-dependent F0 modification in the literature. The evidence we present suggests that the F0 provided by our model is sufficient to generate intelligible and natural sounding utterances, when paired with the utilized neural vocoder. Our findings indicate it is worthwhile to perform a follow-up study, to improve the temporal behavior of F0, e.g., with a different architecture. Also, a self-supervised training scheme may tackle the issue of not having perfect F0 annotations for system training.


\printbibliography
\message{The column width is: \the\columnwidth} 
\end{document}